\documentclass[14pt,preprint,trackchanges]{aastex63}

\usepackage{multirow}
\usepackage{graphicx,epsfig,natbib,color}
\usepackage{epstopdf}
\usepackage{subfigure}
\usepackage{threeparttable}
\usepackage{graphicx}
\usepackage{epsfig}
\usepackage{natbib}
\usepackage{appendix}

\newcommand{\beginsupplement}{%
        \setcounter{table}{0}
        \renewcommand{\thetable}{S\arabic{table}}%
        \setcounter{figure}{0}
        \renewcommand{\thefigure}{S\arabic{figure}}%
     }
     
\received{...} %
\revised{...}
\accepted{...}

\submitjournal{ApJL}

\shorttitle{Mechanism of Microflares}
\shortauthors{Li et al.}

\begin{document}

\title{Three-dimensional Magnetic and Thermodynamic Structures of Solar Microflares}

\correspondingauthor{Xin Cheng and Feng Chen}
\email{xincheng@nju.edu.cn and chenfeng@nju.edu.cn}

\author{Z. F. Li}
\affiliation{School of Astronomy and Space Science, Nanjing University, Nanjing, 210046, People's Republic of China}
\affil{Key Laboratory of Modern Astronomy and Astrophysics (Nanjing University), Ministry of Education, Nanjing 210093, China\\}
\author{X. Cheng}
\affiliation{School of Astronomy and Space Science, Nanjing University, Nanjing, 210046, People's Republic of China}
\affil{Max Planck Institute for Solar System Research, Gottingen, D-37077, Germany}
\affil{Key Laboratory of Modern Astronomy and Astrophysics (Nanjing University), Ministry of Education, Nanjing 210093, China\\}
\author{F. Chen}
\affiliation{School of Astronomy and Space Science, Nanjing University, Nanjing, 210046, People's Republic of China}
\affil{Key Laboratory of Modern Astronomy and Astrophysics (Nanjing University), Ministry of Education, Nanjing 210093, China\\}
\author{J. Chen}
\affiliation{School of Astronomy and Space Science, Nanjing University, Nanjing, 210046, People's Republic of China}
\affil{Key Laboratory of Modern Astronomy and Astrophysics (Nanjing University), Ministry of Education, Nanjing 210093, China\\}
\author{M. D. Ding}
\affiliation{School of Astronomy and Space Science, Nanjing University, Nanjing, 210046, People's Republic of China}
\affil{Key Laboratory of Modern Astronomy and Astrophysics (Nanjing University), Ministry of Education, Nanjing 210093, China\\}

\begin{abstract}
Microflares, one of small-scale solar activities, are believed to be caused by magnetic reconnection. Nevertheless, their three-dimensional (3D) magnetic structures, thermodynamic structures, and physical links to the reconnection have been unclear. In this Letter, based on high-resolution 3D radiative magnetohydrodynamic simulation of the quiet Sun spanning from the upper convection zone to the corona, we investigate 3D magnetic and thermodynamic structures of three homologous microflares. It is found that they originate from localized hot plasma embedded in the chromospheric environment at the height of 2--10 Mm above the photosphere and last for 3--10 minutes with released magnetic energy in the range of $10^{27}-10^{28}$ erg. The heated plasma is almost co-spatial with the regions where the heating rate per particle is maximal. The 3D velocity field reveals a pair of converging flows with velocities of tens of km s$^{-1}$ toward and outflows with velocities of about 100 km s$^{-1}$ moving away from the hot plasma. These features support that magnetic reconnection plays a critical role in heating the localized chromospheric plasma to coronal temperature, giving rise to observed microflares. The magnetic topology analysis further discloses that the reconnection region is located near quasi-separators where both current density and squashing factors are maximal although the specific topology may vary from tether-cutting to fan-spine-like structure.
\end{abstract}

\keywords{Solar flares (1496); Solar magnetic reconnection(1504); Solar coronal heating (1989)}
\section{Introduction}\label{intro}

Solar flares refer to rapid enhancements of electromagnetic emissions in the solar atmosphere and generally originate from strong magnetic field concentrated active regions. In the standard flare model, i.e., so-called CSHKP model (\citealt{1964NASSP..50..451C,1966Natur.211..695S, 1974SoPh...34..323H,1976SoPh...50...85K}), magnetic reconnection is believed to play a substantial role in quickly releasing prestored magnetic energy to heat the plasma. In addition, solar flares are also accompanied with coronal mass ejections (CMEs) \citep[e.g.,][]{Zhang_2001,Cheng_2020} and their association rate usually increases with the magnitude of flares \citep{2005JGRA..11012S05Y}. The rate even reaches 100\% for flares with a peak of soft X-ray (SXR) 1--8 \AA\ flux larger than 3$\times$10$^{-4}$ W m$^{-2}$. 

Differing from energetic flares, microflares are small-scale and short-lived solar activities. They are frequently observed in the regions where the magnetic field is diffuser and weaker than that of active regions (e.g., \citealt{Kuhar_2018}). The energy distribution from microflares to major flares is well known to obey a power-law form (Figure 14 of \citet{2002ApJ...572.1048A}). Although the released energy during once microflare is small, the number is much larger than that of major flares. This means that the total energy deposited into the corona by these small-scale flares is even higher than that by large-scale flares, even may be responsible for corona heating \citep{1991SoPh..133..357H}. Microflares are also found to be closely associated with other small-scale activities such as UV bursts \citep{2014Sci...346C.315P}, coronal bright points \citep[CBPs,][]{2001SoPh..198..347Z}, and hot loops in moss regions \citep[e.g.,][]{2019ApJ...880L..12G,Testa_2020}. These phenomena can be best discriminated and studied by extreme-ultraviolet (EUV) imaging and spectroscopic data. Various observational properties including their morphologies, evolution of magnetic field, and temperature structures suggest that microflares are produced by magnetic reconnection (e.g., \citealt{Qiu_2004,Ning_2008,2010ApJ...712L.111J,2017ApJ...845..122G}), which is also verified by some numerical simulations (e.g., \citealt{Jiang_2010,2012ApJ...751..152J,Archontis_2014}). 

\begin{figure}[!ht]
\centering
\includegraphics[width=18cm]{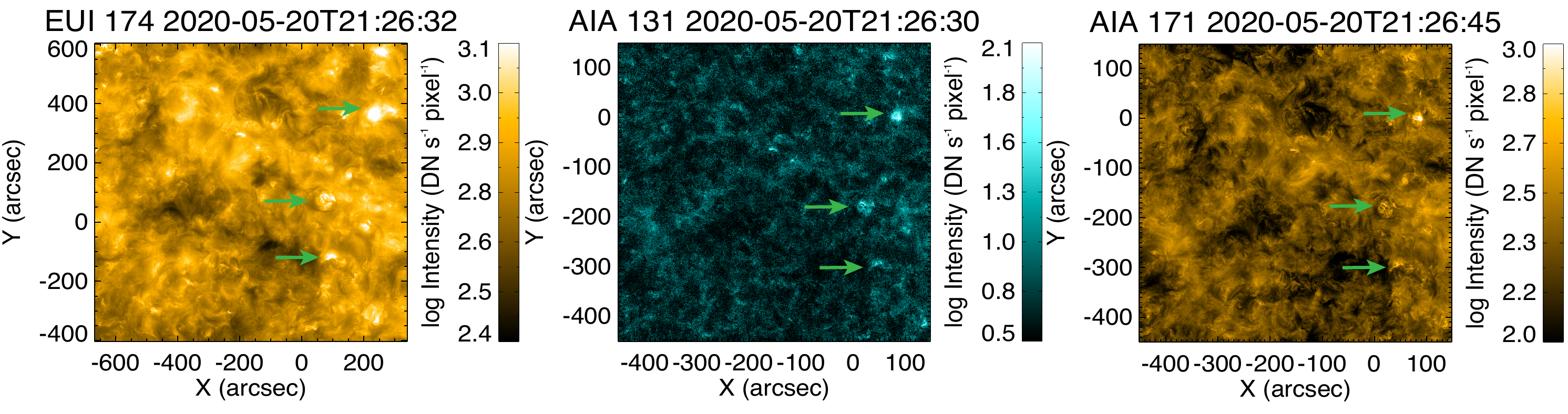}
\caption{EUV images showing three microflares observed on 2020 May 20. The left panel is the EUI/HRI 174 Å image. The middle and right panels display the AIA 131 Å and 171 Å images, respectively. The locations of the three microflares are indicated by three arrows in green.}
\label{fig1}
\end{figure}

Very recently, the Extreme Ultraviolet Imager (EUI; \citealt{2020A&A...642A...8R}) on board Solar Orbiter (SO; \citealt{2020A&A...642A...1M}) performed observations of the quiescent corona in the 174 Å passband near the disk center, which is dominated by \ion{Fe}{9} and \ion{Fe}{10} emissions at 1 MK. The unprecedented high-resolution data collected by High Resolution Imagers (HRIs) of the EUI make it possible to resolve microflares, which was difficult for previous instruments.
Figure \ref{fig1}a shows one snapshot of the EUI 174 Å passband on 2020 May 20 when the Solar Orbiter was located at 0.612 AU away from the Sun. At this moment, the pixel size is approximately 217 km. Figure \ref{fig1}b and \ref{fig1}c display the same field of view but at the 131 Å and 171 Å passbands of the Atmospheric Imaging Assembly (AIA; \citealt{2012SoPh..275...17L}) on board Solar Dynamical Observatory. One can clearly see microflares at different scales at the three EUV passbands but with a higher clarity for the EUI than for the AIA \citep{2021A&A...656L..13C,2021arXiv210410940C}. Besides, the Spectrometer/Telescope for Imaging X-rays (STIX; \citealt{2020A&A...642A..15K}) on board SO also observed many microflares \citep[e.g.,][]{2022arXiv220100712S}.

Nevertheless, only relying on multiwavelength imaging and spectroscopic data, the 3D magnetic and thermodynamic structures behind microflares are still difficult to be determined. To address this question, in this Letter, we mainly focus on high-resolution simulation data with a grid size of 192 km, which is comparable with the spatial resolution of the EUI/HRI on 2020 May 20. In Section \ref{simulation}, we give a brief description of 3D radiative magnetohydrodynamic simulation data we used. Section \ref{results} presents the main results, including 3D magnetic and thermodynamic structures of microflares. Summary and discussion are given in Section \ref{discussion}.

\section{Numerical Simulation Data}
\label{simulation}

The numerical simulation is conducted with the MURaM radiative MHD code (\citealt{2005A&A...429..335V,2017ApJ...834...10R}) that solves fully compressible MHD equations with radiative transfer for optically thick radiation, optically thin radiative loss, and anisotropic thermal conduction. The computational domain covers an area of 197$^2$ Mm$^2$ in the horizontal directions. The vertical extent is 131 Mm with the bottom boundary placed at 18 Mm beneath the photosphere. The domain is resolved by 1024 $\times$ 1024 $\times$ 2048 grid points, corresponding to spatial resolutions of 192 and 64 km in the horizontal and vertical directions, respectively. This simulation is the ``QS run" in a comprehensive simulation of magnetic flux emergence from the convection zone to the corona \citep{2021arXiv210614055C}. More details on the simulation setup and general properties of the coronal plasma and magnetic field are presented in the reference. In brief, the magnetic field in the quiet Sun is a small-scale mixed polarity field maintained by a small-scale dynamo in the convective layers of the domain. The surface magneto-convection provides an upward Poynting flux that is dissipated in the corona. A hot corona of about 1 MK is self-consistently maintained for 11.5 hours evolution of the simulation.

\section{Results}
\label{results}

\subsection{Overview of Microflares}

\begin{figure}[!ht]
\centering
\includegraphics[width=19cm]{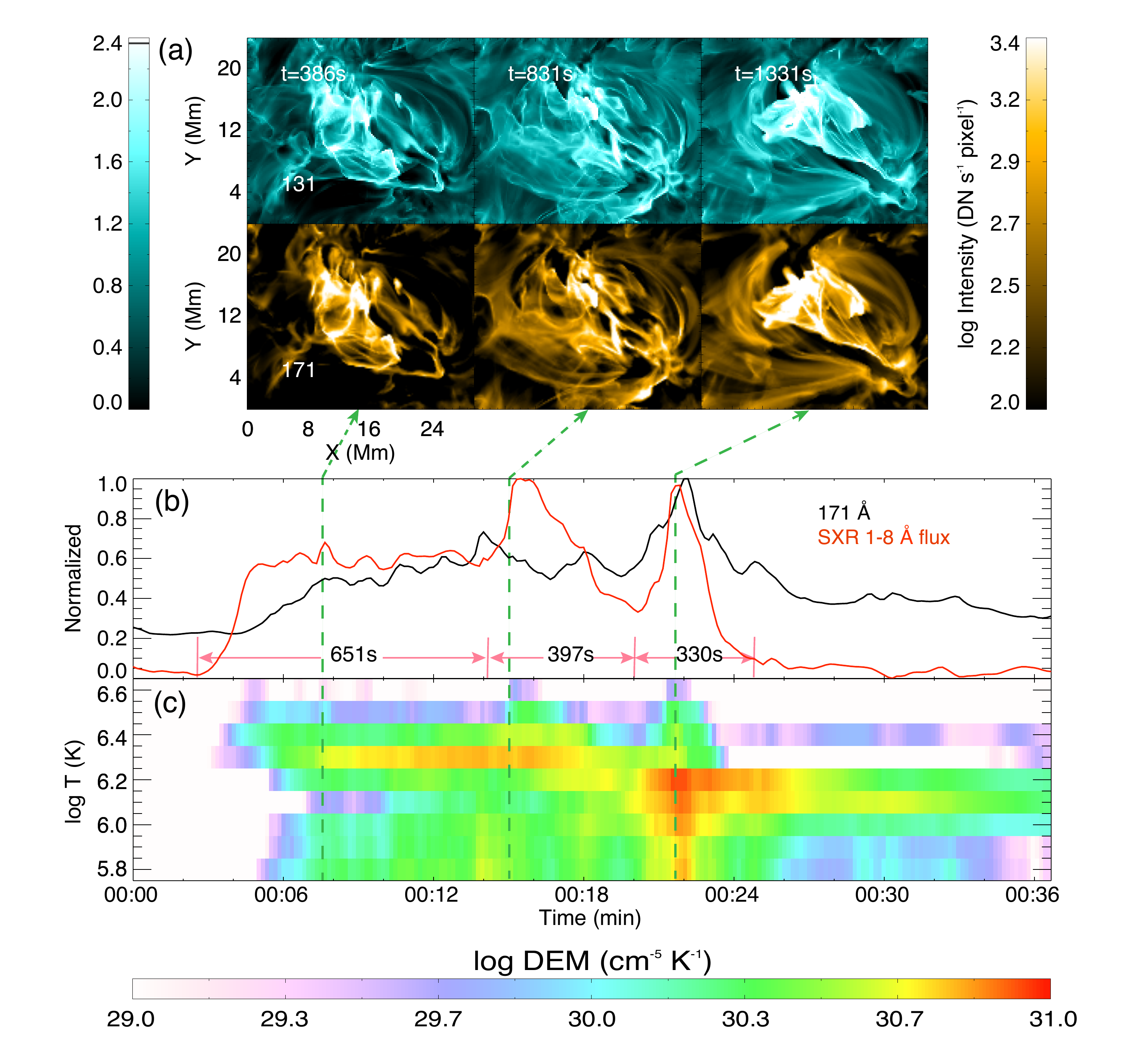}
\caption{Overview of simulated microflares. (a) Snapshots of synthetic 131 \AA\ and 171 Å images at three selected times. (b) Temporal evolution of the SXR 1-8 Å flux (red) and integrated 171 Å intensity (black). Three red line segments indicate three microflares defined by the SXR flux. (c) Temporal evolution of the DEM distribution for the entire region of interest. To address the variations, we subtract the DEM distribution at t$_0$. An animation of this figure is available online. The video shows the evolution of three microflares in 131 Å and 171 Å passbands with a duration of 27 s.}
\label{fig2}
\end{figure}

Taking advantage of the simulated temperature and emission measure, we calculate the total soft X-ray (SXR) 1-8 \AA\ flux and synthesize the AIA EUV images with the AIA response functions \textbf{using the $aia\_get\_response.pro$ procedure, available in the SolarSoft Ware (SSW, \citealt{Freeland:1998we})}. 
Figure \ref{fig2}a shows the synthesized AIA 131 \AA\ and 171 \AA\ images at the three selected times (near the peak times), from which one can see that the simulated microflares present more details than observations. Moreover, the SXR 1--8 \AA\ flux shows that there seems to occur three microflares in succession during the time period of about half an hour with the first one being much longer-lasted than the latter two (Figure \ref{fig2}b; also see the online animation). 

The three homologous microflares originate from a small region of 29 $\times$ 23 Mm$^2$, and are all visible at six synthetic AIA EUV passbands. For each event, its structure is complex and composed of many interlaced loops as shown in Figure \ref{fig2}a. Because of the similarity of the morphology at different passbands, we here only display the AIA 131 \AA\ and 171 \AA\ images. The visibility of the microflares at all AIA EUV passbands indicates that they are multithermal as suggested by previous observations \citep{2014ApJ...789..116I}. \textbf{This property is also confirmed by the differential emission measure (DEM) of the brightenings which distributes in a wide temperature range.} Moreover, the coronal emissions of the three microflares are found to significantly vary with time, indicating a time-dependent energy release rate during these events.

The SXR flux of the three microflares reaches a peak of 10$^{-10}$ W m$^{-2}$, two orders smaller than an A-class flare (Figure \ref{fig2}b). On the basis of the synthetic SXR curve, we define roughly the onset and end times of the three microflares. 
We find that the first microflare has a long duration of about 651 s and the released magnetic energy is $10^{27}$--$10^{28}$ erg. The interesting thing is that the SXR flux seems not to decay after reaching the maximal, with only small fluctuations until the second microflare starts. The second and third microflares are very similar to each other, both presenting an impulsive phase followed by a decay phase, which is consistent with the general evolution pattern for major flares. Their durations are 397 s and 330 s, respectively, and the corresponding magnetic energy release is $10^{27}$ erg, which is basically in the microflare range. 
For all events, the total thermal energies are $10^{28}$--$10^{29}$ erg, consistent with the estimation of microflares recently observed in \citet{2022arXiv220100712S}. However, they also found some non-thermal emission during microflares, which is unrealizable in our MHD simulation.

In order to compare with the SXR 1-8 \AA\ flux, we further calculate the EUV intensities of the microflares at different AIA passbands, which are derived through integrating the entire coronal domain. Figure \ref{fig2}b only shows the result for the AIA 171 \AA\ passband. It is clear to see that, for the first and the third microflare, the evolution of the integrated AIA 171 \AA\ intensity resembles that of the SXR flux. However, for the second one, the variation trend of the 171 \AA\ intensity in time is opposite to that of the SXR flux; the rise (decay) phase of the SXR flux approximately corresponds to the decay (rise) phase of the 171 \AA\ intensity. This is mainly due to the different microflare has distinct temperature structures and evolution patterns, which are further revealed by the evolution of their DEM (Figure \ref{fig2}c). For the first microflare, the DEMs at different temperature bins all present a gradual increase, in particular for the warm plasma in the temperature range of 5.8 $\leq$ log $T \leq$ 6.2. It causes that the enhancement of the 171 \AA\ intensity is more gradual than that of the SXR flux, similar to the events observed by \citet{2022arXiv220100712S}. For the second microflare, the primary emission is contributed by the hot plasma in the temperature range of 6.3 $\leq$ log $T \leq$ 6.5; while, for the plasma in the temperature range of 5.8 $\leq$ log $T \leq$ 6.2, the DEM does decrease, thus giving rise to the decrease of the 171 \AA\ intensity. For the third microflare, the DEM at all temperatures presents an impulsive increase and then a slow decrease, in good agreement with the evolution of the AIA 171 \AA\ intensity and SXR flux.

\subsection{3D Magnetic and Thermodynamic Structures}
\label{3d}

\begin{figure}
\centering
\includegraphics[width=18cm]{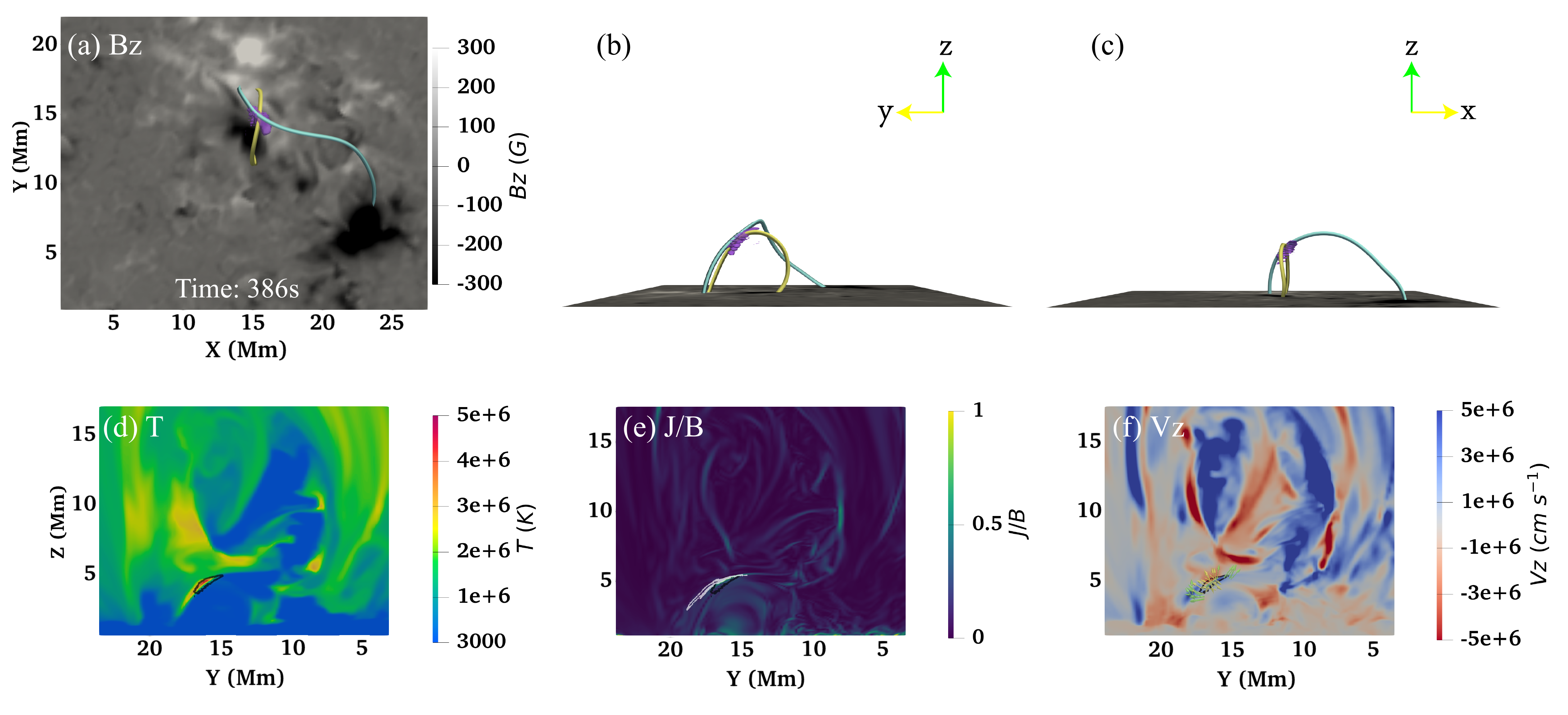}
\caption{3D magnetic structure and reconnection characteristics near the peak time of the first gradual microflare. (a) Photospheric magnetogram overplotted with reconnecting field lines (yellow and blue). The region with a high heating rate is indicated by transparent purple. (b) and (c) Same as (a) but for different perspectives. (d) Temperature distribution in the z-y plane crossing the high heating rate region. The contour in black denotes a high heating rate at a level of 0.5 erg cm$^{-3}$ s$^{-1}$. (e) Distribution of current density normalized by magnetic field J/B in the same z-y plane. The white contour represents the log Q of 1.5. (f) Vertical velocity in the same z-y plane with arrows showing 3D flow field near the reconnection region. The arrows in green (yellow) indicate the outflows (inflows). An animation for the evolution of the temperature and current distributions is available online. The animation proceeds from 386 s to 530 s. The video duration is 2 s.}
\label{fig3}
\end{figure}

\begin{figure}
\centering
\includegraphics[width=18cm]{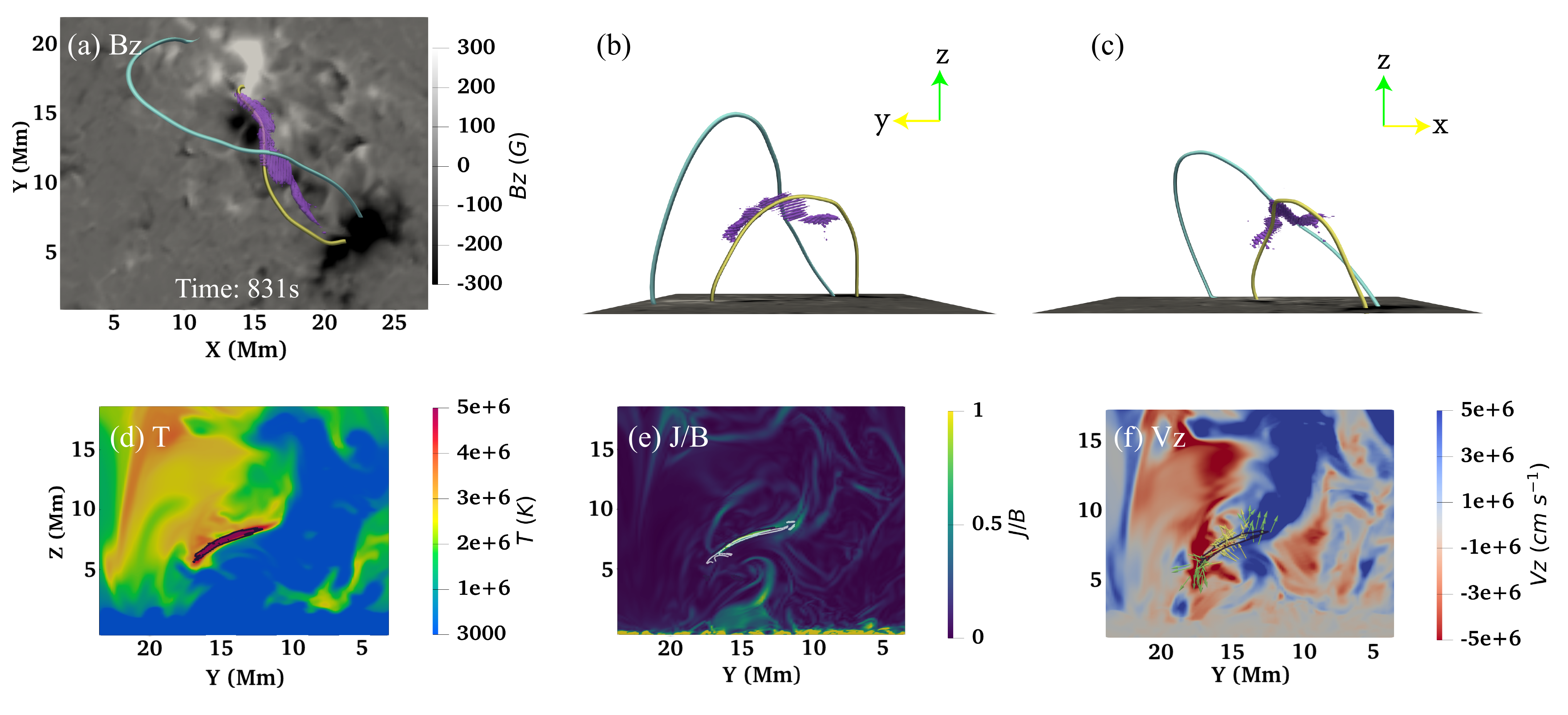}
\caption{Same as Figure \ref{fig3} but for the second impulsive microflare. An animation for the evolution of the temperature and current distributions is available online. The animation proceeds from 779 s to 1176 s. The video duration is 4 s.}
\label{fig4}
\end{figure}

\begin{figure}
\centering
\includegraphics[width=18cm]{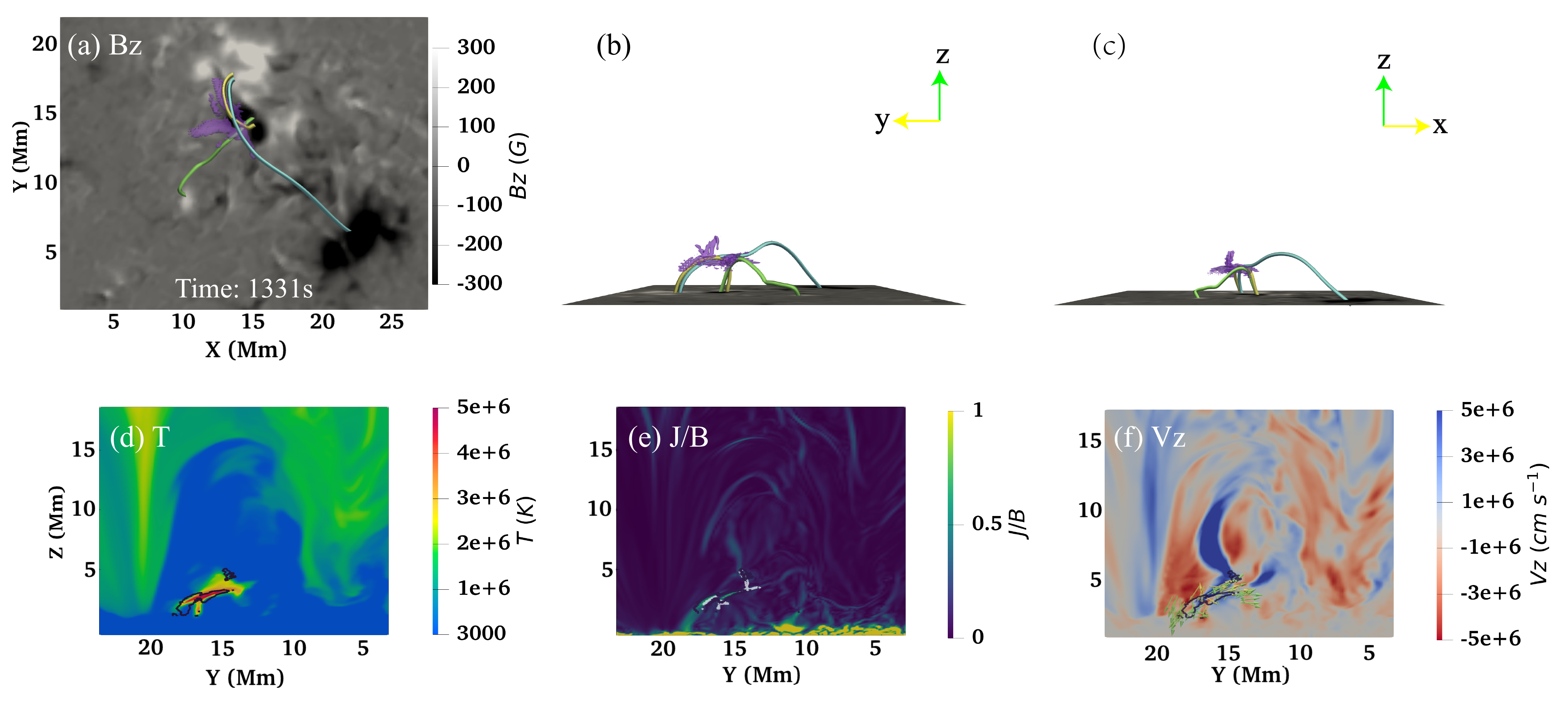}
\caption{Same as Figure \ref{fig3} but for the third impulsive microflare. \textbf{In panel (a), three sets of reconnecting arcades are distinguished with different colors (yellow, blue and green).} An animation for the evolution of the temperature and current distributions is available online. The animation proceeds from 1176 s to 1506 s. The video duration is 3 s.}
\label{fig5}
\end{figure}

To disclose how these microflares are generated and how associated plasma is heated, we analyze the three-dimensional magnetic structures of their source region near the peak times.
Figures \ref{fig3}a displays the vertical magnetic field distribution at the photosphere with an average strength of ~200 Gauss at the base of the first microflare. This value is similar to the strength of the magnetic field at network boundaries (e.g., \citealt{2019LRSP...16....1B}).
It also shows that the microflare is located above a mixed-polarity region that consists of two concentrated polarities. The microflare occurs in the region where a small negative polarity is approaching the major positive polarity.
To locate the energy release of the microflare, we overplot the locations of the strong heating rate per particle, which includes the resistive and viscous terms. It is found that the heated plasma is almost co-spatial with the region where the heating rate per particle is maximal (purple isosurface in Figure \ref{fig3}a). We then trace the magnetic field lines from the regions of high heating rate (Figure \ref{fig3}a-c). It is found that the magnetic configuration of the first microflare primarily consists of two groups of highly sheared arcades. The reconnection most likely takes place at the cross point of them. The heated regions present a curved morphology, indicating that the reconnection site is actually a 3D structure. At the same time, the plasma near the reconnection site is heated and then gives rise to the X-ray and EUV emissions we observed.

Interestingly, we find that the reconnection takes place in the chromospheric environment. Figure \ref{fig3}d shows the temperature distribution at the z-y plane passing the reconnection region. It clearly shows that the temperature around the reconnection region is highly inhomogeneous and the heating mainly occurs at a height of \textbf{4$\sim$5 Mm}, where the plasma is heated up to above 1 MK, almost two orders of magnitude higher than the temperature of the surrounding cool plasma of 10 kK. The process was also found to occur in UV bursts, however, the corresponding temperature and energy are smaller than that for microflares studied here \citep{Hansteen_2017}.
It is also found that the density of the reconnection region ($\sim$ 10$^{11}$ cm$^{-3}$), and the energy release rate is estimated to be approximately 0.5 erg cm$^{-3}$ s$^{-1}$ with a peak value of 5 erg cm$^{-3}$ s$^{-1}$, which is enough to heat the local chromospheric plasma to coronal temperature. This value is even comparable to the heating rate required for major flares \citep{2012ApJ...752..124Q,Liu_2013}, implying that major flares may be composed of plentiful microflares. It needs to be mentioned that the chromosphere in the simulation is treated by assuming local thermal equilibrium, the plasma is thus easier to be heated to a higher temperature \citep{2020A&A...633A..66N}.

The high heating rate is almost cospatial with the strong current, as seen in the current density distribution of the same z-y plane (Figure \ref{fig3}e). The squashing factor Q, which quantitatively describes the degree of change in the connectivity of the field lines \citep{1995JGR...10023443P,2002JGRA..107.1164T} and is calculated through the method developed by \citet{2016ApJ...818..148L}, is shown by the contours in white. It discloses that the reconnection region has the largest Q values. This is in good agreement with the argument that magnetic reconnection is prone to occur in the regions where the magnetic field connectivity changes dramatically and the current density maximizes.

During the reconnection, both reconnection inflows and outflows are observed as displayed by the vertical velocity field at the z-y plane and 3D velocity field around the reconnection region (Figure \ref{fig3}f). The velocities of converging inflows are about $\pm$(30--50) km s$^{-1}$. The velocities of outflows are found to be of nearly 100 km s$^{-1}$. These speeds are comparable with that of reconnection outflows revealed both in observations \citep{2014Sci...346C.315P,2016ApJ...820L..17H} and simulations \citep{2010A&A...510A.111D} in the chromosphere. Note that the velocity for upward outflows is larger than that for downward outflows (Figure \ref{fig3}f),
similar to what was found in a simulation of UV bursts \citep[e.g.,][]{2019A&A...626A..33H}. Taking the Alfv$\acute{e}$n speed at the boundaries of the inflows, the reconnection rate is estimated to be 0.01 $\sim$ 0.1. We also inspect the plasma beta $\beta$, i.e., the ratio of thermal to magnetic pressure, around the reconnection region and find that it is mostly less than 1 (see Table \ref{tab:para}), ranging from 0.001 to 0.01. This shows a high similarity to the reconnection process in the corona as often observed during the CME/flare eruption \citep{2011SSRv..159..421L,2017ApJ...843...21L}.

Because of the long duration of the first microflare, we further inspect the temporal evolution of the 3D magnetic configuration and find that the first microflare primarily consists of three reconnection episodes whose locations are very close to each other. They last for 258 s, 144 s, and 249 s, respectively. The peak of the first microflare actually corresponds to the second reconnection episode. Supplementary Figures \ref{app1} and \ref{app2} show the 3D magnetic field and thermodynamic properties for the first and third episode. One can see that the magnetic field configuration of the reconnection also appears as a tether-cutting type, very similar to that for the second episode. However, the temperature structure seems to change obviously although the reconnection regions are still located in the chromospheric environment. The basic properties of the reconnection regions for the three episodes are summarized in Table \ref{tab:para}.

Figure \ref{fig4} shows the 3D magnetic structures and reconnection characteristics for the second microflare. 
Similar to the first long-duration microflare, the reconnection takes place between two sets of highly sheared arcades but with one set of arcades much longer than the other. The reconnection region also has large current densities and Q values (Figure \ref{fig4}e), as well as surrounded by simultaneous inflows and outflows (Figure \ref{fig4}f). 
While, the difference is that the corresponding reconnection site is located higher, at 5--10 Mm. The plasma density here is about 10$^{10}$ cm$^{-3}$, indicating the environment of the transition region. This could be the reason why the plasma for the second microflare is heated to a higher temperature (almost 5 MK; Figure \ref{fig4}d). \textbf{Moreover, only one reconnection episode occurs during this microflare and effectively heats the plasma. The reconnection region is long elongated (Figure \ref{fig4}a-c) and the magnetic configuration does not vary significantly with time.}

For the third impulsive microflare, it is also caused by \textbf{one reconnection episode} within an elongated current sheet. Similar to the first event, the reconnection takes place in the chromospheric environment but its height (2--3 Mm) is lower than the first two microflares. \textbf{It seems that much energy is released during this event, resulting in the chromospheric plasma near the reconnection region being significantly heated to more than 4 MK.} Note that there are abundant cool plasmas (10 kK) above the reconnection region, thus a fraction of the emission may be absorbed and become invisible. However, the absorption process is not considered when synthesizing EUV images.
The morphologies of the high current density and Q values (Figure \ref{fig5}e) highly resemble that of heated plasma (Figure \ref{fig5}d). The reconnection inflows are mainly contributed by a pair of sheared flows at both sides of the current sheet (Figure \ref{fig5}f). Nevertheless, the 3D structure of the reconnection region is quite different from that for the first and second microflare. As shown in Figure \ref{fig5}a-c, it is more like a 3D fan-spine topology, where three groups of field lines rooted in different polarities concentrate toward the reconnection region, \textbf{and persists throughout the event.}

\begin{deluxetable*}{c|c|c|c|c|c|c|c|c}
\tablenum{1}
\tablecaption{\textbf{Properties of the reconnection during three microflares.}} \label{tab:para}
\tablewidth{0pt}
\tablehead{
\multicolumn{2}{c|}{} & \multicolumn{2}{c|}{Duration (s)} & \multirow{1}*{Height (Mm)} & \multirow{1}*{Size (Mm)} & \multirow{1}*{Temperature (MK)$^{\rm a}$} & \multirow{1}*{Plasma $\beta$} & \multirow{1}*{Topology}}
\startdata
\multicolumn{1}{c|}{\multirow{3}{*}{Event 1$^{\rm b}$}} & Episode 1 & \multicolumn{1}{c|}{\multirow{3}{*}{651}} & 258 & 5 & 6 & 2.5 & 0.002 & Tether-cutting\\ \cline{2-2} \cline{4-9}
\multicolumn{1}{c|}{} & Episode 2 & \multicolumn{1}{c|}{}                     & 144 & 4 $\sim$ 5 & 2.5 & 2 & 0.006 & Tether-cutting\\ \cline{2-2} \cline{4-9}
\multicolumn{1}{c|}{} & Episode 3 & \multicolumn{1}{c|}{}                     & 249 & 5 & 3.6 & 2 & 0.002 & Tether-cutting\\
 \cline{1-9}
\multicolumn{2}{c|}{Event 2} & \multicolumn{2}{c|}{397} & \multirow{1}{*}{5 $\sim$ 10} & \multirow{1}{*}{11} & \multirow{1}{*}{2} & \multirow{1}{*}{0.004} & \multirow{1}{*}{Tether-cutting}\\
   \cline{1-9}
\multicolumn{2}{c|}{Event 3} & \multicolumn{2}{c|}{330} & \multirow{1}{*}{2 $\sim$ 3} & \multirow{1}{*}{4} & \multirow{1}{*}{1.6}& \multirow{1}{*}{0.01} & \multirow{1}{*}{Fan-spine-like}\\
\enddata 
\tablecomments{$^{\rm a}$ Temperature at the DEM peak.}
\tablecomments{$^{\rm b}$ The peak of Event 1 corresponds to Episode 2.}
\end{deluxetable*}

\section{Summary and Discussions}
\label{discussion}

In this Letter, we explore the 3D magnetic structures and thermodynamics of three microflares using RMHD simulation data of a solar quiescent region. The main results are summarized as follows.

1. The microflares are produced by magnetic reconnection that occurs at the regions where the current density and Q values are extremely high, and the reconnection properties are summarized in Table \ref{tab:para}.

2. The reconnection tends to occur in the lower solar atmosphere where the localized plasma is heated from 10 kK to the coronal temperature above 1 MK and then generates EUV emission. For the second microflare, the released energy is even transported to and thus heats the upper corona.

3. The reconnection region appears as a 3D structure, most likely the quasi-separator near the legs of magnetic loops, varing from tether-cutting to fan-spine-like structure for different events.

The energy release mechanism of the three simulated microflares is similar to that for major flares. In the standard CME/flare model, the magnetic energy is believed to be efficiently released through magnetic reconnection, where the current density and Q values are large \citep[e.g.,][]{Sui_2003, 2021RSPSA.47700949L}. Moreover, the reconnection can drive two oppositely directed outflows along the orientation roughly perpendicular to the converging inflows. These features are all observed during the current three microflares. However, there exists a major difference between the large and small-scaled flares, i.e, whether a CME eruption is involved or not. On the one hand, because of no CME eruption, only a small amount of sheared arcades reconnect, thus leading to microflares being short-lived. It is different from the positive feedback process during the CME/flare eruption, in which a number of arcades are involved in the reconnection process. The reconnection accelerates the CME eruption, which in turn further strengthens the reconnection, thus producing a long-duration flare (e.g., \citealt{2000JGR...105.2375L,2016AN....337.1002V,2018ApJ...868..107V}). On the other hand, during the CME/flare eruption, the reconnection configuration will evolve from a hyperbolic flux tube (HFT) to a vertically stretched sheet-like structure \citep{Cheng_2018}. But, during microflares, the reconnection configuration seems not to be deformed significantly during one event and only likely changes from case to case.

Due to the high-resolution simulated data, the first long-duration microflare is found to be composed of three impulsive reconnection episodes. They share very similar magnetic configurations and thermodynamics, even essentially similar to the following two microflares. Nevertheless, since the three episodes in the first microflare occur continuously in close locations, they are hard to be distinguished, in particular, from the evolution of time-integrated observables such as the SXR 1-8 \AA\ flux.

The 3D magnetic and thermodynamic structures of the microflares revealed here shed some lights on understanding the origin of campfires observed by Solar Orbiter. Recently, \citet{2021arXiv210410940C} run a 3D RMHD model for campfires and proposed that they are likely caused by component magnetic reconnection. In their study, the 3D magnetic field structures of the component reconnection are very similar to the tether-cutting type in our study. Observationally, such tether-cutting-like structure was also confirmed to interpret other small-scale transient brightenings similar to campfires \citep[e.g.,][]{2021A&A...656L..13C,2021arXiv211108106M}. Moreover, we also estimate the heights of the microflares, which range from 2 Mm to 10 Mm above the photosphere, similar to that for campfires \citep{2021arXiv210403382B}. However, all previous studies believed that campfires were generated in the corona, rather than in the lower atmosphere as we disclose. Furthermore, \citet{2021arXiv210403382B} pointed out that the reconnection responsible for campfires appear at the apexes of intersected loops; nevertheless, our results show that it more likely occurs at the legs of loops. \textbf{Very recently, \citet{{2022arXiv220306161T}} even found some very tiny dot-like EUV brightenings, part of which were identified to be caused by reconnection at a lower height. In short,} we support the argument that, despite a shorter duration, the campfires belong to the flares family and that the corresponding scale resides between microflares and nanoflares.

It is worth mentioning that the magnetic configuration of the microflares is not always the tether-cutting type. Our results show that it may change to a fan-spine-like structure during the third microflare. In fact, such a fan-spine configuration was also observed in smaller-scale (E)UV bursts \citep{2017A&A...605A..49C}. Generally speaking, the reconnection tends to occur at null point (\citealt{2009PhPl...16l2101P}), separator (\citealt{2005ApJ...630..596L,2010JGRA..115.2102P}) or quasi-separator (\citealt{2005A&A...444..961A}), where the magnetic connectivity changes drastically. After carefully inspecting the magnetic topology of the three microflares, we do not find any signatures of null point and separator. We thus suggest that the quasi-separator reconnection, appearing as tether-cutting (as shown in \citealt{2015NatCo...6.7598S}) or fan-spine-like type, could be more common for small-scale events, at least for the microflares we study. No matter which configuration, it is not always the inverse Y-shaped configuration formed by the reconnection between emerging flux and oblique open flux in the microflare models of \citet{1992PASJ...44L.173S} and \citet{Moore_2010}.

\acknowledgments
We thank the referee who raised valuable comments to improve the manuscript. We also thank Hardi Peter, Pradeep Chitta, and Jie Hong for their helpful discussions. AIA data are courtesy of NASA/SDO, which is a mission of NASA’s Living With a Star Program. Solar Orbiter is a mission of international cooperation between ESA and NASA, operated by ESA. The EUI instrument was built by CSL, IAS, MPS, MSSL/UCL, PMOD/WRC, ROB, LCF/IO with funding from the Belgian Federal Science Policy Office (BELSPO/PRODEX PEA 4000112292); the Centre National d’Etudes Spatiales (CNES); the UK Space Agency (UKSA); the Bundesministerium f$\ddot{u}$r Wirtschaft und Energie (BMWi) through the Deutsches Zentrum f$\ddot{u}$r Luft- und Raumfahrt (DLR); and the Swiss Space Office (SSO). Z.F.L., X.C., and M.D.D. are funded by NSFC grants 11722325, 11733003, 11790303, 11790300, and by National Key R\&D Program of China under grant 2021YFA1600504. F.C. is funded by the Fundamental Research Funds for the Central Universities under grant 0201-14380041. Part of the work is supported by the National Center for Atmospheric Research, which is a major facility sponsored by the National Science Foundation under Cooperative Agreement No. 1852977. The high-performance computing support is provided by Cheyenne (doi:10.5065/D6RX99HX).

\bibliographystyle{aasjournal}

\section{Appendix}
\beginsupplement

\begin{figure}[!ht]
\centering
\includegraphics[width=18cm]{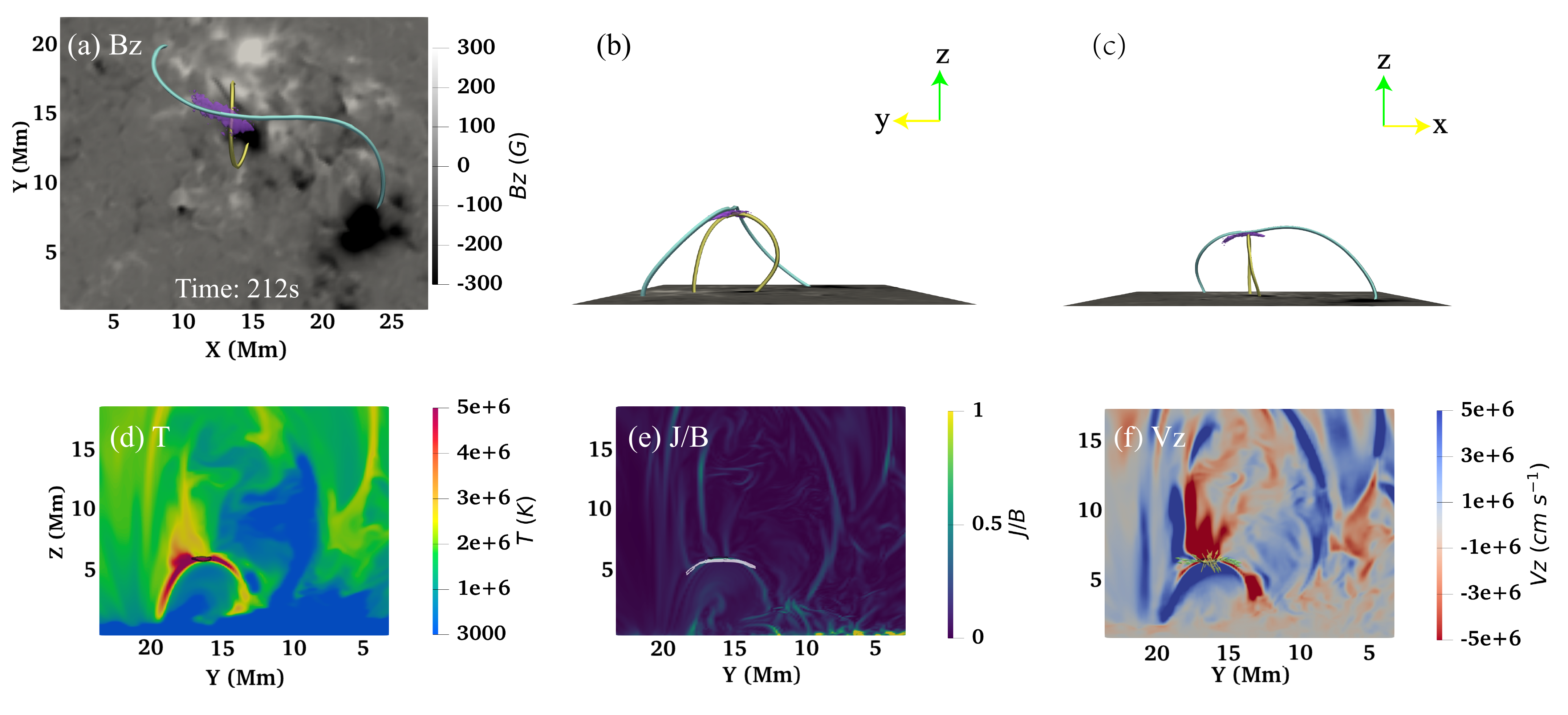}
\caption{Same as Figure \ref{fig3} but for the first reconnection episode of the first microflare. An animation for the temperature and current distributions is available online. The animation proceeds from 128 s to 386 s. The video duration is 3 s.}
\label{app1}
\end{figure}

\begin{figure}[!ht]
\centering
\includegraphics[width=18cm]{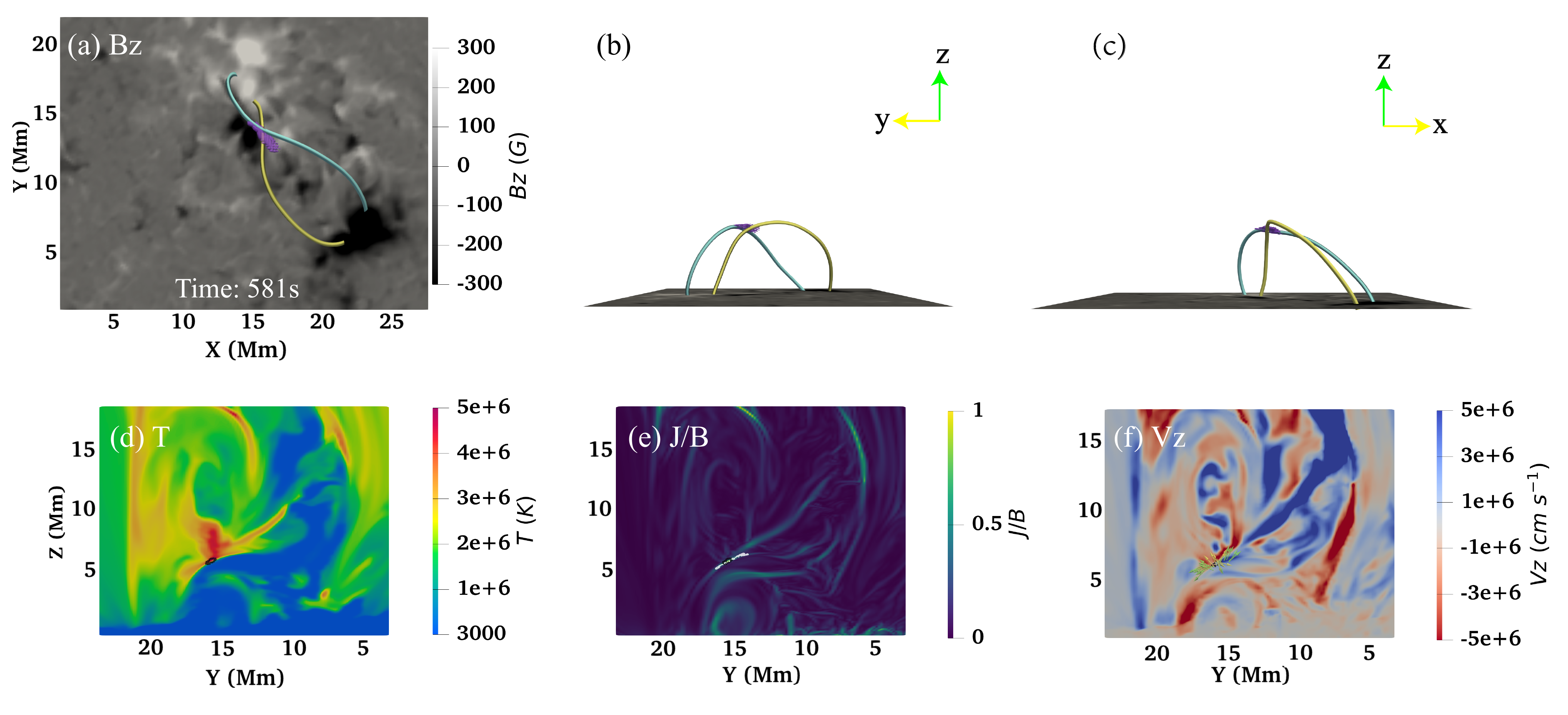}
\caption{Same as Figure \ref{fig3} but for the third reconnection episode of the first microflare. An animation for the temperature and current distributions is available online. The animation proceeds from 530 s to 779 s. The video duration is 3 s.}
\label{app2}
\end{figure}

We find that the first microflare is primarily composed of three reconnection episodes. In Sect. \ref{3d}, we only show the second episode that occurs near the peak time of the SXR 1--8 \AA\ flux. Here, we display the other two episodes as shown in Figure \ref{app1} and Figure \ref{app2}. One can see that, for both of them, the reconnection also takes place between two groups of highly sheared arcades. The main difference is that the reconnection region for the first episode is long stretched, rather than localized during the second episode. Moreover, the peak temperature considerably increases compared with the other two.

\end{document}